\newcommand{\ie}{\emph{i.e.,}\xspace}
\newcommand{\eg}{\emph{e.g.,}\xspace}
\newcommand{\etal}{\emph{et al.}\xspace}
\newcommand{\cut}[1]{}

\newcommand{\squishlist}{
   \begin{list}{$\bullet$}
    { \setlength{\itemsep}{0pt}      \setlength{\parsep}{3pt}
      \setlength{\topsep}{3pt}       \setlength{\partopsep}{0pt}
      \setlength{\leftmargin}{1.0em} \setlength{\labelwidth}{1em}
      \setlength{\labelsep}{0.5em} } }
\newcommand{\squishend}{
    \end{list}  }
    
\newcommand{\squishlistindent}{
   \begin{list}{$\bullet$}
    { \setlength{\itemsep}{0pt}      \setlength{\parsep}{3pt}
      \setlength{\topsep}{3pt}       \setlength{\partopsep}{0pt}
      \setlength{\leftmargin}{2.0em} \setlength{\labelwidth}{1em}
      \setlength{\labelsep}{0.5em} } }
\newcommand{\squishendindent}{
    \end{list}  }

\documentclass[letterpaper,twocolumn,10pt]{article}
\usepackage{usenix,epsfig,endnotes}
\usepackage{listings}
\usepackage{xcolor}
\usepackage[english]{babel}
\usepackage{color}
\usepackage{xspace}
\usepackage[inline]{enumitem}
\usepackage{url}
\usepackage{multirow}
\usepackage{booktabs}
\usepackage{longtable}
\usepackage{afterpage}
\usepackage[flushleft]{threeparttable}
\usepackage[]{hyperref}
\hypersetup{
  colorlinks,
  linkcolor={green!80!black},
  citecolor={red!70!black},
  urlcolor={blue!70!black}
}

\definecolor{mygreen}{rgb}{0,0.6,0}
\definecolor{mygray}{rgb}{0.5,0.5,0.5}
\definecolor{mymauve}{rgb}{0.58,0,0.82}

\newcommand{\projName}{SyzScope}

\newcommand{\numOfHighRiskBugs}{183}

\newcommand{\numOfFixedBugs}{3,396}

\newcommand{\numOfMemoryWriteBugs}{173}
\newcommand{\numOfControlFlowHijackingBugs}{42}
\newcommand{\numOfBugToHighRisk}{183}

\newcommand{\numOfBugOntesting}{1,170}

\begin{document}

\date{}

\title{\Large \bf \projName: Revealing High-Risk Security Impacts of Fuzzer-Exposed Bugs in Linux kernel}

\author{
{\rm Xiaochen Zou}\\
UC Riverside
\and
{\rm Guoren Li}\\
UC Riverside
\and
{\rm Weiteng Chen}\\
UC Riverside
\and
{\rm Hang Zhang}\\
UC Riverside
\and
{\rm Zhiyun Qian}\\
UC Riverside
}

\maketitle

\begin{abstract}
Fuzzing has become one of the most effective bug finding approach for software.
In recent years, 24*7 continuous fuzzing platforms have emerged to test critical pieces of software, \eg Linux kernel.
Though capable of discovering many bugs and providing reproducers (\eg proof-of-concepts), 
a major problem is that they neglect a critical function that should have been built-in,
\ie evaluation of a bug's security impact. 
It is well-known that the lack of understanding of security impact can lead to delayed bug fixes as well as patch propagation.
In this paper, we develop \projName{}, a system that can automatically uncover new ``high-risk'' impacts given a bug with seemingly ``low-risk'' impacts.
From analyzing over a thousand low-risk bugs on syzbot,
\projName{} successfully determined that \numOfBugToHighRisk{} low-risk bugs  (more than 15\%) in fact contain high-risk impacts, \eg control flow hijack and arbitrary memory write,
some of which still do not have patches available yet.

\end{abstract}

\section{Introduction}

Fuzzing is one of the most prolific approaches to discover bugs in software. 
Nowadays, it is even becoming an integral part of the software development process. 
For example, OSS-Fuzz~\cite{oss-fuzz} is a popular platform that can fuzz open-source software continuously.
In addition to OSS-Fuzz which targets primarily user-space applications,
there is also an continuous fuzz testing platform called syzbot~\cite{syzbot} dedicated to fuzz testing Linux kernels. 
The platform has been operational since 2017. Equipped with the state-of-the-art kernel fuzzer Syzkaller~\cite{syzkaller},
it has already reported more than 4,000 bugs to date.

Despite the huge success, such continuous fuzzing platform leads to a major challenge --- the rate of bug discovery 
is much higher than the rate of bug fixes. Taking the syzbot platform as an example, at the time of writing (Jun 2021),
we find that it takes on average 51 days to fix a bug (over \numOfFixedBugs \ fixed bugs), whereas it takes less than 0.4 day for syzbot to report a new bug.

Another critical challenge is the patch propagation to downstream kernels~\cite{E-Fiber}, \eg various PC distributions such as Ubuntu, and the smartphone OS - Android.
Even if the syzbot bugs are patched in Linux, there is often a lack of knowledge on which patched bugs are security-critical. This can have a significant impact on patch propagation delays~\cite{E-Fiber}.
One prominent example is CVE-2019-2215~\cite{cve-2019-2215}, a use-after-free bug that can root most Android devices (\eg Samsung Galaxy S9).
It was initially reported by syzbot and fixed in Linux upstream in 52 days~\cite{fix_cve-2019-2215}.
Unfortunately, it took over a year for the patch to propagate to downstream Android kernels due to the lack of knowledge on its security impact.
In fact, it was only until after bad actors were caught exploiting this vulnerability in the wild did Google start to realize its severity and obtained a CVE number~\cite{cve-2019-2215}.

The previous two challenges illustrate a critical deficiency in today's continuous fuzzing platforms --- lack of automated bug triage, or bug impact analysis.
The goal of this paper is to bridge this gap in the context of Linux kernel bugs.
Performing automated bug impact analyses is challenging and an active area of research, especially in the kernel space. 
Recently, Wu~\etal proposed a solution called \emph{sid} on inferring the security impact of a patch statically~\cite{DBLP:conf/ndss/WuHML20}, \eg use-after-free, or out-of-bound memory access. However, due to the nature of the analysis being completely static, it has to make tradeoffs between soundness and completeness. 
Furthermore, without an actual reproducer, it falls short in determining whether the bug is actually triggerable and exploitable in reality.
On the other end of the spectrum, there has been recent progress on automated exploit generation against kernel bugs, fully demonstrating the exploitability of a bug  by turning a reproducer into an actual exploit.
Fuze~\cite{DBLP:conf/uss/WuCXXGZ18} and KOOBE~\cite{DBLP:conf/uss/ChenZLQ20} are two representative studies that target 
use-after-free (UAF) and out-of-bound (OOB) memory access bugs respectively. 

However, fuzzer-generated bug reports often do not contain the real security impact.
For example, as we find in our experiments, a WARNING error can in fact lead to UAF or OOB.
Furthermore, while UAF and OOB bug types are arguably the most impactful bug types that lead to most real-world exploits, not all of them are created equal. For example, a write primitive, as opposed to read, is always more dangerous than a read, potentially causing corruption of critical pieces of data (e.g., function pointer), leading to arbitrary code execution and privilege escalation. Similarly, a function pointer dereference primitive is also more dangerous than a read primitive.
We define the high-risk and low-risk impacts in \S\ref{sec:definition}.

As evidence that such a distinction between high-risk and low-risk bugs already exists, 
we find, from syzbot's historical data, that UAF/OOB write bugs are often fixed much sooner than UAF/OOB read bugs --- 37 days vs. 63 days for UAF and 29 days vs. 89 days for OOB in terms of average attempt-to-fix delay.
In addition, we also look at the patch propagation delays from upstream to downstream using Ubuntu Bionic as an example.
Patches for WARNING errors have an average delay of 83 days vs. 59 in the case of patches for OOB write bugs.

In lieu of the above, the goal of this paper is to \emph{check whether any of the seemingly low-risk bugs (\eg the ones with read primitives or simply an assertion error) can be turned into high-risk (\eg write primitive or function pointer dereference?})
To this end, we design a system called \projName{} that takes a reproducer and the corresponding bug report as input and produces a report on
any high-risk impacts that are associated with the bug (and optionally new PoCs that exhibits such high-risk impacts).  \projName{} has two distinct modes of operations with respect to the two challenges aforementioned (details are in \S\ref{sec:design}). First, to evaluate the security impact of open bugs and facilitate prioritized bug fixing, we perform a combination of static analysis and symbolic execution to look beyond the initial bug impact reported by the reproducer. Second, to evaluate the security impact of fixed bugs and facilitate timely patch propagation, we additionally allow a fuzzing component as we can use the patch to confirm whether new impacts belong to the same bug.

Surprisingly, after analyzing thousands of seemingly low-risk bugs published on syzbot, \projName{} found \numOfBugToHighRisk{} bugs have high-risk impacts. 
Along the process, we have identified the limitations of the current bug reports and believe \projName{} is a great complement to recognize the hidden impacts of a bug.

In summary, this paper makes the following contributions:

\squishlist
\item
We propose \projName{}, a system that can automatically evaluate the impact of a given seemingly low-risk bug and uncover its true nature.
The system can be easily integrated into the pipeline of syzbot.
\item
To achieve both accuracy and scalability, we have packaged together fuzzing, static analysis, and symbolic execution together
to achieve the end goals. To facilitate reproduction and future research, we open source the system~\cite{SyzScope_github_repo}.
\item
Our tool successfully converts \numOfBugToHighRisk{} seemingly low-risk bugs on syzbot to high-risk bugs including \numOfControlFlowHijackingBugs{} of them with control flow hijacking primitives and \numOfMemoryWriteBugs{} with write primitives. 
\squishend

\section{Background and Overview}

\subsection{Syzbot and Bug Reporting}
\label{sec:syzbot_bug_reporting}

As mentioned earlier, syzbot is a platform that continuously fuzzes the Linux mainline kernel branches. The kernel version advances
on a daily basis so that syzbot always fuzzes the latest version.
All discovered bugs are not only sent to kernel developers and maintainers but also published on an open dashboard in real time~\cite{syzbot}.
For every bug, it also includes valuable information such as bug reports (\eg call trace, perceived bug impact), valid reproducers, specific kernel source version
where the bug was found, kernel configuration files, and patches (if available).
This is a valuable data source that is also suitable for automated analysis.

\begin{table}[t]
    \resizebox{\columnwidth}{!}{
    \centering
    \begin{tabular}{|c|c|}
    \hline
    Bug type & Bug Impact \\ \hline
    \multirow{3}{*}{Sanitizer: KASAN} & use-after-free (UAF) \\
          & out-of-bounds (OOB) \\
          & double-free \\ \hline
    Sanitizer: KCSAN & data race \\ \hline
    Sanitizer: KMSAN & uninitialized use \\ \hline
    Sanitizer: UBSAN & variety* \\ \hline
    Kernel: WARNING / INFO / BUG & Assertions on any unexpected behaviors \\\hline
    Kernel: GPF & corrupted pointer dereference \\ \hline
    \multicolumn{2}{l}{* UBSAN (Undefined Behavior Sanitizer) can detect a variety of impacts} \\
    \end{tabular}}
    \caption{Main impacts of bugs on syzbot}
    \label{tab:Main impacts of bugs on syzbot}
    
\end{table}

\vspace{0.02in}
\noindent\textbf{Bug detectors.}
Syzbot uses the state-of-the-art kernel fuzzer Syzkaller~\cite{syzkaller} which relies on handcrafted ``templates'' or specifications
that encode various knowledge about syscalls, including their relationships (\eg \texttt{open()} and \texttt{close()}) and the domain of syscall arguments. 
During fuzzing, test cases will be generated according to the templates and mutation on the test cases will occur as well. 
Most importantly, there are two general mechanisms to catch bugs at runtime.
First, it leverages various sanitizers that instrument the kernel code to catch memory corruption bugs, 
including Kernel Address Sanitizer (KASAN) ~\cite{KASAN}, Kernel Concurrency Sanitizer (KCSAN)~\cite{KCSAN}, and Kernel Memory Sanitizer (KMSAN)~\cite{KMSAN}, each capable of catching a certain class of bugs (categorized by their impacts, \eg use-after-free).
Undefined Behavior Sanitizer (UBSAN)~\cite{UBSAN} is a special sanitizer that is recently enabled and can detect a variety of bug impacts~\cite{UBSAN_enable}.
Second, it relies on the kernel itself with its built-in assertions such as BUG and WARNING representing uncategorized errors and unexpected behaviors, as well as exception handling, \eg general protection fault (GPF) due to accessing invalid memory pages. 
Whenever a bug is discovered, it must be caught by one of the mechanisms.
The details are listed in Table~\ref{tab:Main impacts of bugs on syzbot}.
In fact, the title of a bug report is automatically generated according to the detection mechanism and the perceived bug impact.
For example, the bug title ``KASAN: use-after-free Read in hci\_dev\_do\_open'' indicates that it is caught by KASAN and 
the perceived bug impact is use-after-free read.

\vspace{0.02in}
\noindent\textbf{Limitations of the current bug detectors.}
One principle currently embraced by fuzzers is that the execution of some buggy input
stops as soon as any bug impact is discovered. This is because the goal of a fuzzer
is to discover new bugs and fix them. When the first error is caught, it is pointless to let
the program continue executing because it is already in a corrupted state. Any subsequent buggy behaviors 
are further and further away from the root cause of a bug, and therefore do not really contribute in 
understanding and fixing a bug.
However, this principle does not help with realizing the maximum impact of a bug.
In fact, it is the opposite of what we need because a bug can often lead to multiple impacts,
some of which may not be immediately uncovered and may even require additional syscall invocations to manifest.

\subsection{High-Risk vs. Low-Risk Bug Impacts}
\label{sec:definition}

Based on the recent literature~\cite{DBLP:conf/uss/WuCXXGZ18,DBLP:conf/uss/ChenZLQ20,DBLP:conf/ndss/WuHML20, 236346} and the recent high-profile
bugs that are exploited in practice~\cite{cve-2018-9568,cve-2019-2025,cve-2019-2215}, we define high-risk bug impacts to be the following:

1. Any UAF and heap OOB bugs\footnote{Subsequently,
we refer to heap OOB bugs as OOB bugs for brevity} that lead to a \emph{function pointer dereference primitive}, which is effectively a control flow hijacking primitive that can likely lead to an end-to-end exploitation (\ie arbitrary code execution in the kernel context)~\cite{236346}. Such primitives can happen for example when the function pointer is located in a freed object or out-of-bound memory, and is incorrectly dereferenced.

2. Any UAF and OOB bugs that lead to a \emph{write primitive}, including overwriting a freed object or an object out-of-bounds, and in general any writes to unintended locations and/or with unintended values (\eg a write by dereferencing an unsafe data pointer). Write primitives, as opposed to read, have the opportunity to corrupt control data (\eg function pointers) and can be effectively turned into control flow hijacking as well. In addition, write primitives can be used for data-only attacks that achieve privilege escalation without explicitly altering the control flow (\eg by modifying the uid of a process)~\cite{PT-Rand}.

3. Any invalid free bugs. This includes freeing a memory area that should not be freed or an already freed object (the latter corresponds to double-free bugs). Invalid frees can be turned into a UAF of multiple different candidate objects chosen by an adversary~\cite{slake}. With the freedom of choice of various candidate objects, the likelihood of finding a function pointer dereference or write primitive is high.

In contrast, a low-risk impact is defined to be anything other than the above.
This includes any UAF or OOB bugs that lead to only read primitives (without write or function pointer dereference primitives), as well as any other impacts such as WARNING, INFO, BUG, and GPF, as defined in Table~\ref{tab:Main impacts of bugs on syzbot}.
Finally, we will give a more detailed breakdown of the high-risk impacts by their primitives in \S\ref{sec:validation}.

\begin{figure}[t]
    \centering
    \begin{minipage}[t]{0.9\linewidth}
        \begin{lstlisting}[language=C]
static void tcindex_free_perfect_hash(struct tcindex_data *cp) {
  for (int i = 0; i < <@\textcolor{red}{cp->hash}@>; i++)
    <@\textcolor{red}{tcf\_exts\_destroy}@>(<@\textcolor{mygreen}{\&cp->perfect[i].exts}@>);
  kfree(cp->perfect);
}

void tcf_exts_destroy(struct tcf_exts *<@\textcolor{mygreen}{exts}@>) {
  if (<@\textcolor{mygreen}{exts->actions}@>) 
    <@\textcolor{red}{tcf\_action\_destroy}@>(<@\textcolor{mygreen}{exts->actions}@>);
}

int tcf_action_destroy(struct tc_action *<@\textcolor{mygreen}{actions[]}@>) {
  struct tc_action *a;
  for (i = 0; i < TCA_ACT_MAX_PRIO && <@\textcolor{mygreen}{actions[i]}@>; i++) {
    <@\textcolor{mygreen}{a = actions[i];}@>
    <@\textcolor{mygreen}{actions[i] = NULL};@>  // AAW
    ret = <@\textcolor{red}{\_\_tcf\_idr\_release}@>(<@\textcolor{mygreen}{a}@>);
  }
}

int __tcf_idr_release(struct tc_action *<@\textcolor{mygreen}{p}@>) {
  if (<@\textcolor{red}{\_\_tcf\_action\_put}@>(<@\textcolor{mygreen}{p}@>, ...))
    ...
}

static int __tcf_action_put(struct tc_action *<@\textcolor{mygreen}{p}@>, ...) {
 struct tcf_idrinfo *<@\textcolor{mygreen}{idrinfo}@> = <@\textcolor{mygreen}{p->idrinfo}@>;
 if (refcount_dec_and_mutex_lock(&<@\textcolor{mygreen}{p->tcfa\_refcnt}@>, &<@\textcolor{mygreen}{idrinfo->lock}@>)) {
  ...
  <@\textcolor{red}{tcf\_action\_cleanup}@>(<@\textcolor{mygreen}{p}@>);
 }
}

static void tcf_action_cleanup(struct tc_action *<@\textcolor{mygreen}{p}@>) {
  if (<@\textcolor{mygreen}{p->ops->cleanup}@>)
    <@\textcolor{red}{p->ops->cleanup(p);}@>  // FPD
}
        \end{lstlisting}
    \end{minipage}
    \begin{tablenotes}
      \small
      \item AAW = Arbitrary address write
      \item FPD = Function pointer dereference
    \end{tablenotes}
    \caption{A slab-out-of-bounds Read bug on syzbot}
    \label{fig:A example case of a slab-out-of-bounds Read bug}
\end{figure}

\subsection{Motivating Example}

To illustrate why that is the case, we use a real bug from Syzbot~\cite{case_study_3} as an example to demonstrate how it is possible to turn a low-risk \emph{slab-out-of-bounds read} bug into a control flow hijack exploit.

As shown in Figure \ref{fig:A example case of a slab-out-of-bounds Read bug}, the bounds of the iteration \texttt{cp->hash} (line 2) can be turned larger than the size of array \texttt{cp->perfect} (which resides on the heap), creating a potential OOB situation(the vulnerable object marked in green). More precisely, at line 8, an OOB read access occurred via \texttt{exts->actions}, leading to a slab-out-of-bounds read.
Syzkaller stops at line 8 and generates a bug report with the title of ``KASAN: slab-out-of-bounds Read in tcf\_exts\_destroy'' on syzbot.
Interestingly, even if we allow the execution to continue forward during fuzzing, 
it will almost always end up with an exception at line 14 because it is 
highly likely that \texttt{actions[i]} attempts to retrieve an element from an invalid address, \ie it is equivalent of *(actions+i). Note that \texttt{actions} itself can point to any random location because it was read out-of-bounds (see line 8 and 9).
As a result, one may come to the conclusion this is indeed a low-risk bug.

However, since the entire \texttt{exts} structure is out-of-bounds, it is actually possible for an attacker to control the data at an appropriate offset by spraying a number of objects nearby~\cite{DBLP:conf/uss/ChenZLQ20}.
Specifically, if the correct data is sprayed, \texttt{actions[i]} will retrieve an element from a valid address and prevent the kernel from crashing at line 14.
After that, we are able to observe an arbitrary address write opportunity at line 16.
This is because the pointer \texttt{action} comes from the OOB memory \texttt{exts}, 
which means \texttt{action} can potentially point to any arbitrary memory address. 
Even further down,
we can see a control flow hijack opportunity arises at line 36 through a function pointer dereference, where the value of the function pointer \texttt{p} is controlled by an attacker as well (basically \texttt{actions[i]}).
By now, we can conclude that this bug is actually very much high risk and needs to be patched as soon as possible. 
Interestingly, no one seemed to have realized the potential impact of the vulnerability.
As a result, the bug was silently fixed without any CVE being assigned and it took almost 4 months (much longer than the average time-to-fix).
The fact that there is no CVE assigned would also delay downstream kernels from applying the patch.

\begin{figure*}[t]
    \centering
    \includegraphics[width=1\textwidth]{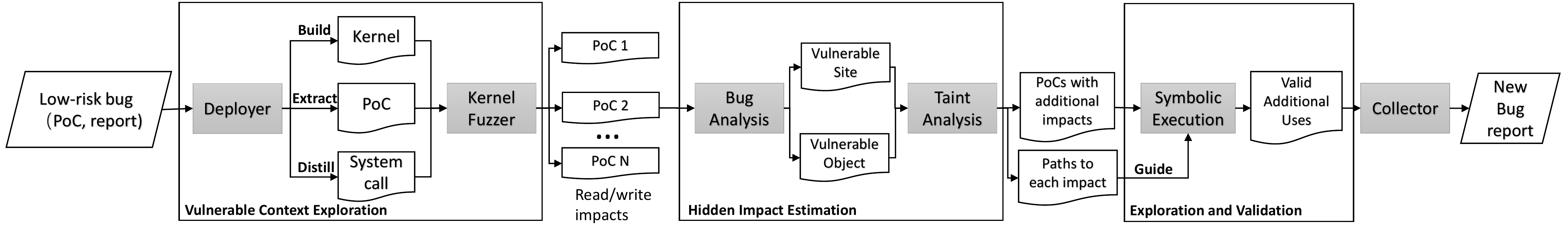}
    \caption{Workflow}
    \label{fig:Workflow}
\end{figure*}

\subsection{Goals and Non-Goals}
\label{sec:scope}

\noindent\textbf{Goals.}
\projName{} aims to reveal high-risk impacts of a seemingly low-risk bug by analyzing subsequent behaviors since the first reported impact and alternative paths where high-risk impacts may be located.
As alluded to earlier, we believe there are two benefits provided by \projName{}.
First, it will facilitate prioritized bug fixing. This is especially important given that the number of kernel bugs that are discovered 
on a daily basis due to continuous fuzzing and the transparency offered by the syzbot platform.
Second, it will speed up the patch propagation in Linux-derived ecosystems. Even when a patch is already available in Linux mainline,
it can take months and even years to propagate to all its downstream kernels, \eg Android~\cite{DBLP:conf/ndss/WuHML20,E-Fiber}.

\vspace{0.015in}
\noindent\textbf{Non-goals.}
\projName{} does NOT aim to produce an end-to-end exploit automatically,
which would require the automation of a number of additional steps such as reliable heap feng shui, bypass of various defenses including KASLR, SMEP, and SMAP~\cite{236346}.
Therefore, we do not claim that the high-risk impacts we defined are 100\% exploitable.
Instead, we \emph{aim to uncover as many such high-risk impacts as possible, within a reasonable time and resource budget}.
The more high-risk impacts, the more likely an exploit can be generated. 
In fact, our work is complementary to the research on automated exploit generation. 
For example, KOOBE~\cite{DBLP:conf/uss/ChenZLQ20} is designed to take an OOB write bug and turn it into a control flow hijack primitive. 
\projName{} can turn a seemingly non-OOB bug or an OOB read into an OOB write such that KOOBE can take it further and prove its actual exploitability. 
Finally, \projName{} does NOT aim to evaluate bugs whose impacts are outside of the types listed in Table~\ref{tab:Main impacts of bugs on syzbot}.

\section{Design}
\label{sec:design}

In this section, we describe the design of \projName{}. 
The intuition driving the design is that uncovering more impacts of a bug is fundamentally a search problem.
Interestingly, even though a fuzzer is designed to essentially search through an input space as well,
its goal is to optimize for maximum code coverage and/or number of bugs; it is never to maximize the impacts of a given bug.
Therefore, \projName{} is designed to perform a much more targeted search starting from a PoC that already uncovers some impact of a bug.

\vspace{0.05in}
\noindent\textbf{High-level Workflow}
The workflow is depicted in Figure~\ref{fig:Workflow}, which contains three main components:
\emph{Vulnerable Contexts Exploration},
\emph{Hidden Impacts Estimation},
and \emph{Validation and Evaluation}.
They correspond to the three main techniques that we leverage and integrate together: fuzzing, static analysis, and dynamic symbolic execution.
At a high level, given a PoC and its bug report that demonstrates some low-risk impact,
we first start with a ``fuzzy'' approach by performing a targeted fuzzing campaign to explore other potential vulnerable contexts, \ie additional impacts.
This will allow us to cast a wider net early on. 
Second, given these additional PoCs and impacts (where some may be high-risk already and may still be low-risk), 
we then leverage static analysis to locate any potential high-risk impacts in alternative execution paths that we have not yet been able to reach during fuzzing. 
Finally, the static analysis will guide the execution of symbolic execution to confirm whether these high-risk impacts are reachable in practice.

\subsection{Vulnerable Contexts Exploration}
\label{sec:design_fuzzing}

\begin{figure}[t]
    \centering
    \begin{minipage}[t]{1\linewidth}
        \begin{lstlisting}[language=C]
void __rxrpc_put_peer(struct rxrpc_peer *peer) {
 struct rxrpc_net *rxnet = peer-><@\textcolor{mygreen}{local->rxnet}@>;
 ...
}

void rxrpc_send_keepalive(struct rxrpc_peer *peer){
 ...
 whdr.epoch=htonl(peer-><@\textcolor{mygreen}{local->rxnet->epoch}@>);
 ...
 peer-><@\textcolor{mygreen}{local->socket->ops->sendmsg()}@>
 ...
}
        \end{lstlisting}
    \end{minipage}
    
    \caption{Impacts of the same bug located in different syscalls}
    \label{fig:Buggy context may have different hidden impacts.}
\end{figure}

As can be seen in Figure~\ref{fig:Workflow}, the \emph{vulnerable context exploration} component takes a seemingly low-risk bug (including a PoC and 
a corresponding bug report), and produces one or more new PoCs that exhibit additional impacts (either low-risk or high-risk).
The problem with the original PoC is that it traverses only a single path in the kernel and can therefore be very limited in terms of impact coverage. By exploring more contexts (\ie paths) associated with the bug, we are more likely to uncover additional impacts.
In general, there are two logical possibilities that additional impacts are located.
First, it may be hidden directly behind the reported impact, \ie in the same invocation of a syscall as shown in Figure~\ref{fig:A example case of a slab-out-of-bounds Read bug}.
Second, it may be triggered in a completely different syscall not present in the original PoC,
and may even require additional syscalls to set up the state beforehand and removal of existing syscalls to undo some state. 
Figure \ref{fig:Buggy context may have different hidden impacts.} illustrates a real case from syzbot (simplified) 
to show two different impacts of the same bug.
In the original PoC, it causes \texttt{peer->local} to be freed accidentally before entering function \texttt{\_\_rxrpc\_put\_peer()},
leading to a UAF read. However, the PoC fails to discover the additional function pointer dereference which is located
in \texttt{rxrpc\_send\_keepalive()}.
In addition, \texttt{rxrpc\_send\_keepalive()} is not a function reachable by \texttt{\_\_rxrpc\_put\_peer()}.
Therefore, a good test case would need to insert an additional syscall (with the right arguments) after the
syscall that triggered the free.

To this end, we will leverage Syzkaller to mutate the original PoC and explore any missed additional impacts.
As mentioned in \S\ref{sec:syzbot_bug_reporting}, the goal of a fuzzer is never to uncover a bug's maximum impact.
However, it is suitable for Syzkaller to search for new contexts that may be locked away in a specific sequence of syscalls~\cite{syzkaller_intro}. 
More specifically, we define \textit{the original context} as the execution path exercised by the original PoC. \textit{Any new context} by definition must be associated with a different execution path that must not share the same initial primitive (UAF/OOB read or write). 
Along with vulnerable context exploration, we also log all the impacts along each path and attribute any high-risk impacts to the corresponding context. 

Nevertheless, there are three challenges in making the search effective. 
First, as illustrated earlier in Figure~\ref{fig:A example case of a slab-out-of-bounds Read bug}, 
an earlier impact can block the execution from exploring deeper parts of the code and uncovering additional impacts. 
Second, even though a coverage-guided fuzzing strategy is approximately aiming to explore more code blocks and paths,
it does not directly recognize the importance of impacts. In fact, it is possible that a new impact is triggered in an already explored path,
\eg OOB access can happen in array element access 
only when an argument (representing the array index) of a syscall becomes large enough. 
Third, we need to ensure that the mutated PoC, should it uncover any new impacts, is still exercising the same bug.
Indeed, if we allow Syzkaller to mutate any part of the PoC, \eg by removing a critical syscall or adding arbitrary syscalls,
then it is entirely possible that the mutated test case will trigger a different bug altogether. 

To overcome the challenges, we make two important changes to Syzkaller and made an important observation.

\vspace{0.015in}
\noindent\textbf{Impact-aware fuzzing.}
To address the first challenge, we will attempt to carry on the fuzzing session even when impacts are detected (by sanitizers or the kernel itself).
Specifically, only a few bugs like \emph{general protection fault} or some \emph{BUG} impacts are irrecoverable errors (\eg a NULL pointer dereference and divide by zero);
other ones can be simply safely ignored. 
For example, when an OOB read impact occurs, we simply allow the kernel to read any data from out-of-bounds memory without panic and continue executing.
This way, it is possible that KASAN catches another OOB write eventually. 
To address the second challenge, we note again that Syzkaller is an entirely coverage-guided fuzzer, \ie its feedback metric is only the coverage.
However, it tends to focus its attention and energy quickly on newly covered code, leaving covered but potentially buggy code space behind.
Therefore, we introduced the feedback of impact.
The idea is that Syzkaller will now consider a test case as a seed not only when it discovers any new coverage but also when it
discovers any new impact. 
As it is rare to discover test cases that uncover new impacts, we always assign higher priority to them for mutation. 

\vspace{0.015in}
\noindent\textbf{Restricted-fuzzing space.}
By default, Syzkaller considers all syscalls in the operating system as candidates to insert into a test case (during mutation).
However, clearly this is undesirable for uncovering impacts of a specific bug.
For instance, a bug in the \texttt{kvm} module should not allow the TCP/IP modules' syscalls, which
will likely drive the test cases to cover code outside of \texttt{kvm}.
Therefore, to address the third challenge, we restrict the syscalls to those Syzkaller templates that subsume the syscalls in the PoC.
In addition, we preserve all the syscalls in the original PoC and allow only insertion of syscalls. 
We also allow the mutation of arguments of existing syscalls.
This strategy aims at preserving the root cause of a bug while still allowing new impacts to be discovered.
If the above restricted-fuzzing strategy cannot uncover any new impacts or coverage (currently after a threshold of 5 minutes),
we will activate another slightly more aggressive fuzzing strategy, where we will relax the restrictions
to allow syscalls against the entire module, \eg all syscalls related to the networking module or kvm module (one module sometimes can correspond to multiple templates). 
In addition, we also allow the removals of syscalls in the original PoC (though the latter is set to a very low probability). 

\vspace{0.015in}
\noindent\textbf{Side effects of fuzzing.}
Unfortunately this does still result in test cases that trigger different bugs.
In fact, as we later find out, close to half of the new impacts are associated with a completely unrelated bug,
where the other half are indeed new impacts belonging to the original bug.
Therefore, we choose to perform fuzzing only when we have a strategy to confirm whether a new impact belongs to the original bug.
Specifically, where patches are already available, \ie the second use case of \projName{} as discussed in \S\ref{sec:scope}, 
we develop a heuristic that can do so accurately. The details are deferred to \S\ref{sec:implementation}.

\subsection{Hidden Impacts Estimation}
\label{sec:design_static}

From the previous step, we have likely found more PoCs that uncovered additional impacts.
However, it is possible that they are still low-risk impacts. 
After all, it is challenging for a fuzzer to explore deeper part of the kernel code, especially when 
there are complex conditions that may prevent it from making progress.
Therefore, we can invoke the hidden impact estimation component to determine whether further high-risk impacts beyond the low-risk ones are available.
Specifically, given an existing low-risk impact, we leverage static analysis to conduct a type-specific search.
At the moment, we support the analysis of two types of low-risk impacts, \ie UAF read and OOB read. Again, we choose them because UAF and OOB are two of the most dangerous bug impact types. In order to discover high-risk impacts such as write or function pointer dereference impact, 
we formulate the problem as a static taint analysis problem as will be articulated later.

This component requires two inputs: a PoC that triggers one or more UAF/OOB read impacts, and the corresponding bug report as input.
It then extracts two pieces of critical information from the report: (1) vulnerable object and (2) vulnerability point(s).
A \emph{vulnerable object} is defined to be the freed object (in the UAF case) and the object intended to be accessed yet an out-of-bound access occurred (in the OOB case).
A \emph{vulnerability point} is the statement where the use of a freed object happens (in the UAF case) or the OOB memory access (in the OOB case) occurs.

We first observe that the UAF/OOB read impacts are due to the read of some data that can be potentially controlled by an adversary, \ie either a freed and refilled object or something sprayed next to the vulnerable object.
This means that any subsequent use of the read memory can be potentially dangerous. 
Specifically, we consider dereferences of function pointers that are derived from such memory high-risk impacts,
\ie UAF/OOB function pointer dereferences (as mentioned in \S\ref{sec:scope}).
Similarly, we consider any dereferences of data pointers that are derived from such memory high-risk impacts 
if they lead to memory writes (to an attacker-controlled address), \eg \texttt{*data\_ptr = 0}.

The second observation is that UAF/OOB write impacts are basically writes to the same planted memory by an adversary.
For example, if a planted object whose function pointer is overwritten due to a UAF/OOB write,
it can potentially lead to arbitrary code execution, as line 32 in the example presented in Figure~\ref{fig:A example case of a slab-out-of-bounds Read bug} shows.

To summarize, the above high-risk impacts can be formulated as a static taint analysis problem.
For pointer dereferences, the taint source will be the \emph{vulnerable object}.
The sink would be the dereference of a pointer (either function pointer or data pointer that leads to a write). 
Whenever any source flows to the sink, we consider it a high-risk impact.
It is slightly different for UAF/OOB write. The taint source is the same. However, we do not need any sinks. 
Instead, whenever any value is written to the tainted memory, we consider it a high-risk impact.
In the example presented earlier in Figure~\ref{fig:A example case of a slab-out-of-bounds Read bug},
line 16 is such a write to the OOB memory. 

In addition to reporting the additional impacts, we also record the branches that definitely need to be taken
and the ones that definitely need to be avoided in order to reach the them. This is later used to guide the symbolic execution.
We defer the implementation details of the static taint analysis to \S\ref{sec:implementation}.

\subsection{Validation and Evaluation}
\label{sec:validation}

As shown in Figure \ref{fig:Workflow}, this module takes a PoC with potentially new high-risk impacts,
aim to achieve two goals in this component:
(1) validating the feasibility of reaching these high-risk impacts,
and (2) evaluating the more fine-grained impact in terms of the enabled primitives, \eg arbitrary value write vs. constrained value write. 

\vspace{0.015in}
\noindent\textbf{Validate the feasibility of high-risk impacts.} 
In this step, we symbolize the same vulnerable object, \ie taint source in static analysis,
dynamically at the point when the vulnerable object is first read, \eg an OOB read or UAF read.
The entire vulnerable object is symbolized with the assumption that an attacker can spray desired payloads onto the heap.
However, a well-known limitation of symbolic execution is its scalability due to path explosion. 
Therefore, we additionally leverage static analysis to guide the symbolic execution in two ways:
(1) When an additional impact is found to be far away(40 basic blocks is our current threshold, see the details at ~\ref{sec:comp_by_comp}) from the initial impact, we take the path guidance from static analysis results
in terms of which branches must and must not be taken. For example, an additional impact can be located in only a specific true branch, and therefore
the false branch does not need to be explored.
This avoid unnecessary explorations during symbolic execution and therefore speed up the process.
Currently we choose to apply the guidance only when the additional high-risk impact is at least 40 basic blocks away intraprocedurally .
This is because if they are nearby, symbolic execution will have no problem locating them anyways. 
Furthermore, we are not able to rule out any potential false negatives from static analysis and therefore it is best to let symbolic execution explore all possible paths.
(2) Based on the farthest potential high-risk impact, we limit the scope of the analysis up to that point. Otherwise, the symbolic execution
may continue executing until the end of a syscall, taking much more significant time.

\vspace{0.015in}
\noindent\textbf{Evaluate the fine-grained impacts.} 
As mentioned earlier in \S\ref{sec:definition},
the high-risk impacts can include three types of primitives: \textit{write}, \textit{func-ptr-deref}, and \textit{double free}.
Below we discuss the \textit{write} and \textit{func-ptr-deref} primitive.

(1) \emph{Overwriting symbolic memory}. 
The impact of such a write requires a more careful analysis. For example, when such an OOB write occurs,
it is necessary to understand the offset of the write, the length of the write, and the data that can be written~\cite{DBLP:conf/uss/ChenZLQ20}. 
Then we can evaluate whether the write is flexible enough to overwrite a function pointer in a heap-sprayed object nearby.
Since such an analysis is already done in \cite{DBLP:conf/uss/ChenZLQ20}, we simply refer to such writes as \emph{UAF write} or \emph{OOB write}
and leave them to existing work for further triage.

(2) \emph{Write with symbolic data or write to symbolic address}. 
When the write target is symbolic or data to-be-written is symbolic, We follow a classic definition of a write primitive is ``write-what-where''~\cite{write-what-where}. 
For the ``what'' dimension, it will be either an ``arbitrary value'' or ``constrained value'' write,
depending on the absence or presence of any symbolic constraints on the value. 
For the ``where'' dimension, it will be either an ``arbitrary address'' or ``constrained address'' write,
again depending on the symbolic constraints of the address.
More concretely, considering a write instruction \texttt{mov qword ptr [rax], rdi}. It attempts to store the data in \texttt{rdi}
to an address stored in \texttt{rax}. If \texttt{rdi} is symbolic, it may be considered an arbitrary value or constrained value write.
Similarly, if \texttt{rax} is symbolic, it may be considered an arbitrary address or constrained address write. Note that \texttt{KASAN} cannot detect such primitives because it is designed to catch the initial read of freed/OOB memory. Subsequent propagation of them via writes may require taint tracking.

(3) \textit{Dereferencing symbolic function pointer} is detectable by monitoring the symbolic status of the function pointer. Similar to the write primitives, \texttt{KASAN} can only detect the initial problematic read, and not the subsequent use (\ie function pointer dereference) of the freed/OOB memory.

(4) \textit{Passing pointer (either being symbolic or pointing to symbolic memory) to free function}. 
To detect invalid frees (including double frees), we examine the pointer argument of heap free functions such as \texttt{kfree()}. 
If the pointer itself is symbolic, it indicates that an attacker can control the memory that will be freed, therefore corresponding to the invalid free cases. If the pointer itself is not symbolic but the memory the pointer points to is, it is considered a double free (we group these two cases together as invalid frees).

\section{Implementation}
\label{sec:implementation}

In this section, we describe details about each component in \projName{}.
In total, the system has over 10K lines of code and is fully automated.

\subsection{Vulnerable Context Exploration}
\label{sec:impl_fuzzing}

There are several things worth describing in more details in this component.

\vspace{0.015in}
\noindent\textbf{Impact-aware fuzzing.}
Syzkaller has two major components: (1) the fuzzer itself that runs on a target OS where test cases are executed (\ie syz-fuzzer),
and (2) a manager that runs outside of the target OS, overseeing the fuzzing process (\ie syz-manager).
The fuzzer is designed to mutate test cases locally and only send test cases that contribute to more coverage back to the manager.
However, different for traditional coverage-guided fuzzing, our impact-aware fuzzing will mutate a PoC that already leads to an impact,
which essentially causes a crash of the entire target OS.
This often leads to a burst of crashes, which means the fuzzer will lose the states of the mutation because of the crash.
By default, the manager will simply log the test case but no further mutation will be performed, 
because it is likely going to trigger the same bug over and over again.
In our implementation, we enable the manager to remember the impact-inducing PoC and send it to the fuzzer for mutation (currently 500 times), 
preempting any other mutation of regular seeds. If a new impact is discovered, the corresponding PoC will then be treated similarly.

\vspace{0.015in}
\noindent\textbf{Multiple impacts in a single PoC.}
As mentioned earlier in \S\ref{sec:syzbot_bug_reporting}, Syzkaller by default allows only the first impact to be reported while ignoring all the rest.
This means that if a PoC happens to trigger multiple bug impacts, \eg one UAF read and another UAF write,
the later impacts are hidden.
This contradicts with our goal of recovering more bug impacts.
Therefore, we turn on a \texttt{KASAN} booting option called \texttt{kasan\_multi\_shot} when booting the kernel, which will present all impacts instead of only the first.
However, there are some impacts that are not possible to ignore, \eg a NULL pointer dereference which would cause an irrecoverable crash.
We can bypass some other assertion-related kernel panics by disabling \texttt{panic\_on\_warn} in the kernel boot options or some options in the kernel config like \texttt{CONFIG\_BUG\_ON\_DATA\_CORRUPTION}.
Note that these debugging options are turned on specifically for fuzzing or debugging (as they are useful in catching errors). 
In practice, they are off by default in production settings.

\label{sec:Confirming the impact belonging to the same bug}
\vspace{0.015in}
\noindent\textbf{Confirming the impact belonging to the same bug.}
As described earlier in \S\ref{sec:design_fuzzing}, when a bug already has a patch available 
(we assume that these patches are correct),
we use a heuristic that can use the patch to test whether a new PoC and its new impacts (generated from fuzzing) are still a result of the same bug.
The idea is as follows.
If a new PoC can still trigger the same impact after the patch (as well as all prior commits) is applied, clearly the new PoC is triggering a different bug.
However, if it no longer works, we are not sure if it is because of the patch itself that breaks the PoC or one of the earlier commits does so.
To deal with this ambiguity, instead of applying the patch and its prior commits up to that point, we will attempt to apply the patch commit itself directly. 
If it can be successfully applied, \ie git does not reject it and kernel compiles/boots normally without breakage, then if the new PoC
no longer works, we say that the PoC is indeed exploiting the same bug.
Otherwise, if it cannot be successfully applied, we will instead apply all the commits up to the patch (but not including the patch itself)
and retest the PoC. If it can reproduce the impacts, then we know that these intermediate commits do not interfere with the bug triggering. 
Therefore, we know that the new PoC is still triggering the same bug, because earlier we have verified that it does not work against a patched kernel (with all prior commits applied also).
The one last remaining corner case is that the PoC cannot reproduce the impacts when we apply all the commits up to the patch (but not including the patch itself).
In such cases, we will act conservatively and simply give up the new PoC because it remains ambiguous.

\subsection{Hidden Impacts Estimation}
\label{sec:impl_static}

Our static analysis engine is built on top of \texttt{DR. CHECKER}~\cite{DBLP:conf/uss/MachirySCSKV17},
which is flow-sensitive, context-sensitive, field-sensitive, and inter-procedural.
We made some changes to adapt it to our scenario as described below. 

\vspace{0.015in}
\noindent\textbf{Interfacing with the fuzzer result.}
Recall that our static analysis engine takes a vulnerable object and vulnerability point(s) as input.
Since DR. CHECKER is based on LLVM IR, the input would also be given at the LLVM IR level.
However, the fuzzing result (\texttt{KASAN} report) includes such information at the binary level,
thus requiring a mapping from the binary to IR.
Specifically, it includes (1) the call trace that contains vulnerability points that trigger the UAF/OOB bug (binary instruction addresses). 
(2) the size of a vulnerable object and the offset at which the vulnerable object is accessed (in either UAF or OOB). 

With such information, we will first try to locate the vulnerable function that triggered the impact in LLVM IR. Usually, the vulnerable function is the first function on 
the call trace besides KASAN-related ones. However, if the vulnerable function is inlined, 
we then resort to its caller (potentially recursively if multiple layers of inlining occur). 
To simplify the handling of inlined functions, we choose to compile the kernel using Clang without any inlining to generate the kernel bitcode. 
After locating the vulnerable function in LLVM IR, we leverage the Clang-generated debug information 
to map each IR instruction back to a corresponding source line. 
Specifically, we look for a \textit{load} IR instruction that maps to the same source line number as what is reported in the 
vulnerable function in the \texttt{KASAN} report.
Once we successfully locate such a \textit{load} instruction that triggers the bug, 
we can retrieve the base pointer (vulnerable object) of the \textit{load} instructions and taint the object.

\vspace{0.015in}
\noindent\textbf{Trace recorder}: 
In order to generate the branch guidance for symbolic execution (described in \S\ref{sec:design_static}), 
we record the detailed taint trace from an initial vulnerability point to the high-impact one, 
including the calling context and instruction of each taint propagation.
This way, if the taint propagation occurs in a specific call sequence and specific branch,
we can precisely drive the dynamic symbolic execution accordingly.

\subsection{Validation and Evaluation}  
\label{sec:impl_validation}

There are two possible symbolic execution engines we can potentially use: S2E~\cite{s2e} and angr~\cite{angr}.
S2E is a great candidate as it is designed to support dynamic (in-vivo) symbolic execution, such that most of the memory locations 
are populated with runtime data. This limits the symbolized memory to a much smaller scope, \ie those that are potentially controlled by an adversary.
Unfortunately, S2E supports only a single CPU core in each QEMU instance~\cite{s2e_single_core}. Therefore, it has a major drawback in its ability to reproduce race condition bugs.
In our preliminary experiments, we find that over half of the race condition bugs simply cannot be reproduced reliably within a reasonable time frame, 
\ie an hour.
Therefore, we decided to first reproduce bugs in a vanilla QEMU. Once the bug is successfully reproduced (\texttt{KASAN} report being triggered).
The breakpoint we set in \texttt{KASAN} report function will immediately freeze the memory of QEMU, and provide the corresponding CPU registers and memory address of the vulnerable object (either freed object or an object that is out-of-bounds) to angr.
Whenever angr needs to access a memory location, we will look up their actual values on the snapshot on demand.

\section{Evaluation}
\label{sec:evaluation}

\subsection{Dataset and Setup}
\label{sec:setup}

We evaluate \projName{} against the majority of low-risk bugs reported on syzbot.
At the time of writing, there are 3,861 low-risk bug reports in total according to our definition.
We exclude the ones detected by KCSAN, KMSAN, UBSAN.
This is because they are either not mature enough yet (KMSAN and UBSAN) and do not contain critical information in the bug reports for us to continue the analysis (\eg vulnerable object), 
or do not have any valid reproducer (none of the KCSAN bugs has a reproducer). 
After this step, there are 3,267 remaining bug reports.

Next, we filter the bug reports that do not target the upstream Linux kernel (which is our main focus)
and those that do not contain any reproducer (either a Syzkaller program or a C program).
Then, we divide the dataset into ``fixed'' and ``open'' sections. 
For the fixed cases, since each bug report comes with a corresponding patch, we can deduplicate bug reports based on shared patches (unfortunately we are unable to do so for the bugs in the open section),
\eg the bug reports may look different but their root causes are the same.
In our tests, we pick only one bug report from the group that share the same patch. Specifically, we use the one with the highest risk impact.
For example, if a low-risk impact bug report (\eg UAF read) happens to share the same root cause (\ie same patch) with a high-risk one (\eg UAF write),
we eliminate the entire group of reports because the corresponding bug should be recognized as high-risk already.
If a WARNING bug report happens to share the same patch with a UAF read report, we will pick the UAF one as input to \projName{},
because it already contains critical information such as the vulnerable object, allowing us to continue the analysis in the pipeline.
Otherwise, we simply randomly pick a bug report among the available ones (WARNING, INFO, BUG, or GPF) as there is no major difference.
Finally, we obtain \numOfBugOntesting{} bug reports (after deduplication) and their corresponding reproducers. 

In our current configuration, to be conservative, 
we do not attempt the vulnerable context exploration step on bug reports 
in the open section since we can not verify the new impacts still belong to the same bug (as discussed in \S\ref{sec:design_fuzzing}).
All experiment are conducted in Ubuntu-18.04 with 1TB memory and Intel(R) Xeon(R) Gold 6248 20 Core CPU @ 2.50GHz * 2.
For each bug report and its reproducer, we allocate a single CPU core for 3 hours of kernel fuzzing maximum, 
1 hour of static analysis (it usually finished within half an hour), and 4 hours of symbolic execution.

\begin{table*}[t]
  \centering
  \resizebox{\textwidth}{!}{
  \begin{tabular}{@{}llllllrrrrrrr@{}}
      \toprule
& \multirow{2}{*}{Initial Bug Impact} & \multirow{2}{4em}{Raw bug reports} & \multirow{2}{4em}{Valid bugs*} & \multirow{2}{4.5em}{High-risk bugs*} & \multirow{2}{4em}{High-risk impacts} & \multicolumn{7}{c}{High-risk impact breakdown by primitive type} \\ \cmidrule{7-13}
&    &  &  &  &   & UOW   & AAW   & CAW   & AVW   & CVW   & FPD & IF\\ \midrule
    \multirow{3}{*}{Fixed}  & GPF and BUG      & 687  & 215  & 17  & 323   & 71     & 124     & 62     & 29     & 20     & 8 & 9\\
                           & WARNING and INFO  & 918  & 293  & 15   & 379   & 85     & 166     & 66     & 20     & 30     & 9 & 3 \\
                         &    UAF and OOB Read & 680  & 202  & 99  & 2866  & 319     & 1490     & 446     & 271     & 153     & 104 & 83\\\midrule
    \multirow{3}{*}{Open}  & GPF and BUG       & 225  & 83   & 4   & 6     & 4     & 0     & 0     & 0     & 0     & 0 & 2\\
                           & WARNING and INFO  & 541  & 292  & 10   & 501   & 97     & 213     & 91     & 47     & 22 & 18  & 13\\
                         &    UAF and OOB Read & 216  & 85   & 38  & 768   & 151     & 381     & 113     & 43     & 22     & 40 & 18\\\midrule
      & Sum                                    & 3267 & 1170 & 183 & 4843  & 727     & 2374     & 778     & 410     & 247 & 179 & 128\\ \bottomrule
    \multicolumn{13}{l}{* Valid bugs and high-risk bugs in the fixed section are all unique ones (after deduplicating based on the bug reports) } \\

  \end{tabular}}
  \begin{tablenotes}
      \small
      \item UOW= UAF/OOB write,  AAW= Arbitrary address write,  CAW= Constrained address write,  AVW= Arbitrary value write 
      \item CVW= Constrained value write,  FPD= Function pointer dereference,  IF= Invalid Free
    \end{tablenotes}
  \caption{Overall Results}
  
  \label{tab:Overall Results}
\end{table*}

\subsection{Overall Results}
\label{sec:overall_results}

In total, out of the \numOfBugOntesting{} low-risk bugs analyzed by \projName,
we report that \numOfBugToHighRisk{} of them turn out to contain at least one high-risk impact.
This is more than 15\% of all the bugs.
Furthermore, out of the \numOfBugToHighRisk{} bugs, 
\numOfMemoryWriteBugs{} have at least one write primitive (\eg UAF/OOB write, arbitrary address write).
\numOfControlFlowHijackingBugs{} of them have at least one func-ptr-defer primitive.

We break the results down, first by open vs. fixed bugs, and then by the initial bug impact type.
The results are shown in Table~\ref{tab:Overall Results}.
In addition, we categorize the high-risk impacts into 7 primitives, as defined in \S\ref{sec:validation}
and listed in the table.

First, comparing the results for open and fixed bugs, we can see in general there is a higher percentage of low-risk bugs turned into high-risk in the fixed section compared to the ones in the open section (18.4\% vs. 11.3\%). 
In addition, the average number of primitives for each bug in the fixed section is 27.2 versus 24.5 in the open section.
This is because the open bugs did not go through the vulnerable context exploration phase,
due to the concern that newly discovered contexts may belong to different bugs (as mentioned in \S\ref{sec:design_fuzzing}).
As another evidence demonstrating the utility of fuzzing, we find that bugs in the fixed section have 1.33 contexts on average compared to the only 1 context per bug in the open section. Since each context has about 22 primitives on average, it is clear that fuzzing will allow more primitives to be uncovered.
Nevertheless, even without the help of fuzzing, \projName{} still managed to turn 52 open bugs from low-risk into high-risk.

In addition, it is worth noting that bugs with initial impacts being UAF/OOB read have a much higher chance
of turning into high-risk cases. In the fixed section, 99 out of 202 UAF/OOB read cases can be turned into high-risk.
Furthermore, the total number of high-risk impacts for just the 99 cases is 2866 (on average 29 per case).
In the open section, 38 out of 85 UAF/read cases (slightly smaller fraction) can be turned into high-risk.

For GPF, BUG, WARNING, and INFO, they usually represent a diverse set of root causes
which are masked by the kernel itself or simply exceptions. In reality, there are still 
a subset of these cases that are really memory corruption bugs.
Indeed, we are able to discover 46 high-risk cases out of 883 in these categories across fixed and open sections.
In general, GPF and BUG cases typically represent more serious bugs in the kernel (compared to WARNING and INFO) 
and some of these impacts can lead to exceptions that are difficult to bypass (as mentioned previously, \eg NULL pointer dereference). 
As a result, we find that the number of new impacts discovered under them is very small in the open section,
as we did not perform any fuzzing.

With respect to the impact primitives, 
We can see that the arbitrary address write has by far the highest number. 
This demonstrates the necessity of applying symbolic execution, as such impacts can only be uncovered
with such analyses.
If one can write to an arbitrary address, it is typically a strong primitive (even if the write value is not arbitrary) 
that is highly exploitable, especially when combined with a read primitive~\cite{androidbinder}.

The second most common types of primitives are arbitrary value and constrained value write (778 and 410 cases in total), 
followed by UAF and OOB wirte (727 in total).
Writing arbitrary values can be dangerous but its exploitability depends highly on where we can write to. 
The UAF and OOB write are generally serious as an attacker can often choose multiple different objects (e.g., containing a function pointer) to overwrite~\cite{DBLP:conf/uss/ChenZLQ20,DBLP:conf/uss/WuCXXGZ18}.

The func-ptr-defer primitives are relatively rare (179 cases in total) but highly exploitable~\cite{236346}.
We sampled a few cases for further analysis
and managed to write three working PoCs that can successfully hijack the control flow of the kernel. 
One of them will be described in the case studies later.
Finally, we find 128 invalid free primitives, which can be turned into adjustable UAF bugs as mentioned in \S\ref{sec:definition}. %

\label{sec:comp_by_comp}
\subsection{Component by Component Analysis}

\begin{table}[t]
    \centering
    \setlength\tabcolsep{1pt}
    \resizebox{\columnwidth}{!}{
    \begin{tabular}{@{}lcccl@{}}
    \hline
    \multirow{2}{*}{Initial bug impact}  & \multirow{2}{2.3cm}{High-risk bugs by fuzzing}  &  \multirow{2}{3cm}{Primitives found by fuzzing} & \multirow{2}{2.4cm}{Extra high-risk bugs by S\&S} & \multirow{2}{2.4cm}{Extra primitives found by S\&S}\\ 
    & & & & \\ \hline
    GPF and BUG         & 13          & 37   & 4 & 285 \\ 
    WARNING and INFO  &   11      & 33      & 4  & 346            \\ 
    UAF and OOB Read  &   42      & 128     & 57 & 2738 \\ \hline
    \end{tabular}}
    \begin{tablenotes}
      \small
      \item S\&S = Static analysis and symbolic execution
    \end{tablenotes}
    \caption{Result breakdown by components on fixed bugs}
    \label{tab:Results in detail}
\end{table}

Now that we have shown the end-to-end results, we break them down by components to understand the contribution of each.
Here we focus on fixed bugs only because we applied all the components in \projName{}.

Specifically, we show the intermediate results obtained from fuzzing alone, and then the additional results from performing the static analysis and symbolic execution (S\&S) on top of fuzzing. 
As a main result shown in Table~\ref{tab:Results in detail},
we can see that 66 bugs are turned from low-risk to high-risk by fuzzing alone,
and an extra 65 bugs are turned after the static analysis and symbolic execution (S\&S). 

\vspace{0.015in}
\noindent\textbf{Vulnerable Context Exploration.} 
Even though fuzzing is generally effective, 
due to the lack of systematic path exploration as is done by symbolic execution,
we see that the number of primitives found by fuzzing alone is limited (as shown in Table~\ref{tab:Results in detail}).
This is because every context (i.e., path) provided by fuzzing is concrete and thus only a limited number of primitives may be covered. Furthermore, as mentioned in \S\ref{sec:validation}, fuzzing relies on KASAN which by design is unable to recognize the more indirect types of primitives such as arbitrary address write or control flow hijacking.
This is the reason why we still continue with S\&S even if the bug is determined to be high-risk by fuzzing.
For example, we may find a UAF write primitive through fuzzing, but through S\&S we may find an even more serious control flow hijacking primitive. 
Nevertheless, new contexts can be beneficial for S\&S, as they can be considered as ''seeds'' fed into the S\&S to explore many more paths and uncover more primitives. 
As we mentioned in \S\ref{sec:overall_results}, there are on average 1.33 contexts on average per fixed bug from fuzzing.

On a different note, we verified the heuristic we proposed in \S\ref{sec:impl_fuzzing} to confirm that the new impacts found through fuzzing 
indeed belong to the same bugs.
To do so, we manually analyzed 10 sampled new impacts corresponding to 10 different bugs found by \projName{}.
We are able to confirm they are all caused by the same bug.

\vspace{0.015in}
\noindent\textbf{Hidden Impact Estimation.}
Here, we evaluate the effectiveness of using static analysis to guide the symbolic execution. 
In particular, we sampled 53 bugs and evaluated them with and without the guidance using static analysis. 
We observed 16 bugs, with guidance, experienced a 12x-190x speedup compared to no guidance,
which allowed the symbolic execution to finish in minutes as opposed to hours.
In addition, we found 2 bugs whose high-risk impacts can be found only with guided symbolic execution, and the unguided version simply finds nothing in four hours.

However, we note that there is a fundamental tradeoff between running time and false negatives.
This is because our static analysis is not sound and can potentially miss important primitives.
Therefore, as mentioned in \S\ref{sec:validation}, we set a threshold of 40 basic blocks and only primitives beyond the distance of 40 will trigger the guidance. 
We also evaluate the choice of the threshold by varying it as 30, 40, and 50. 
With a threshold of 30, we observe seven more false negative cases compared to the threshold of 40, due to the fact the static analysis missed those primitives that are located in a distance between 30 and 40. 
With a threshold of 50, we find two primitives in a distance between 40 and 50 will take more than four hours to finish (as they become unguided).
Note that this threshold is considered suitable for our experiment setup only.
With more resources to expend on symbolic execution, 
the threshold can be further increased.

\vspace{0.015in}
\noindent\textbf{Exploration and Validation.} 
As we mentioned earlier, on top of the 66 bugs that were turned from low-risk to high-risk by fuzzing alone, \projName{} turned an additional 65 bugs (via S\&S) from low-risk into high-risk, \ie from zero primitive to at least one. 
The result shows that the different components are complementary to each other. 
Note that even if fuzzing has already found a primitive, we still applied the static analysis and symbolic execution to find even more primitives, to obtain a more complete picture of the bug impact.
In fact, the majority of the high-risk primitives are attributed to S\&S, as shown in Table~\ref{tab:Results in detail}.

We also considered the scenario where we omit the fuzzing phase altogether and apply S\&S directly to the original context provided by the reproducer. Interestingly, we find that it is still capable of turning all but two bugs into high-risk (albeit with a a subset of primitives discovered).
This shows that there are many bugs that had at least one primitive reachable from the original context.

\subsection{False Positives \& False Negatives}
\label{sec:FPFN}
\textbf{False Positives.}
By design, \projName{} confirms all the high-risk impacts dynamically through either fuzzing or dynamic symbolic execution, and therefore should not incur any false positive.
Nevertheless, in practice, we do make implementation-level simplifications for scalability considerations that can potentially lead to false positives. 
For example, during the dynamic symbolic execution, we skip a list of common kernel functions (a total of 51) such as \texttt{printk(), kasan\_slab\_free()} for performance reasons, which can potentially lead to unwanted side effects. Fortunately, we have not observed any false positives because of this.

\vspace{0.015in}
\noindent\textbf{False Negatives.}
It is expected that \projName{} will incur false negatives as the solution is opportunistic in nature. 
To understand the extent of the problem, we manually inspected 83 bugs and noticed that the false negatives can come from three different sources.
\textit{(1) Failing to find more vulnerable contexts.} We observed 4 bugs with an incomplete set of primitives because \projName{} fails to find the appropriate vulnerable contexts. 
Specifically, race condition is the major reason in such cases.
\textit{(2) Imprecise static analysis.} We observed 2 bugs that have an incomplete set of primitives because of the imprecise result from static analysis. 
\textit{(3) Timeout of symbolic execution.} We observe 3 bugs that have an incomplete set of primitives
due to the early termination of symbolic execution (currently a 4-hour timeout). We discover these cases by increasing the timeout to 16 hours. 
In summary, all of the above false negatives can be potentially reduced when given more computational resources, \eg more fuzzing time and symbolic execution time (without having to rely on the precision of static analysis). 
We also note these 9 bugs are all turned into high-risk already because \projName{} found other primitives.

\subsection{Case Studies}

\begin{figure}[t]
    \centering
    \begin{minipage}[t]{0.9\linewidth}
        \begin{lstlisting}[language=C]
static int bfs_create(...){
 ...
 ino = <@\textcolor{red}{find\_first\_zero\_bit}@>(<@\textcolor{mygreen}{info->si\_imap}@>, info->si_lasti + 1);
 if (ino > info->si_lasti) {
  ...
  return -ENOSPC;
 }
 <@\textcolor{red}{set\_bit}@>(ino, <@\textcolor{mygreen}{info->si\_imap}@>);
 ...
}

unsigned long find_first_zero_bit(const unsigned long *<@\textcolor{mygreen}{addr}@>, unsigned long size)
{
 unsigned long idx;
 for (idx = 0; ...) {
  if (<@\textcolor{mygreen}{addr[idx]}@> != ~0UL)
   return min(idx * BITS_PER_LONG + ffz(<@\textcolor{mygreen}{addr[idx]}@>), size);
 }

 return size;
}
\end{lstlisting}
    \end{minipage}
    
    \caption{Code snippet of a arbitrary address write}
    \label{fig:Code snippet of a arbitrary address write}
\end{figure}

\begin{figure}[t]
    \centering
    \begin{minipage}[t]{0.9\linewidth}
        \begin{lstlisting}[language=C]
static void tcp_check_sack_reordering(struct sock *sk, ...)
{
 struct tcp_sock *<@\textcolor{mygreen}{tp}@> = tcp_sk(sk);
 ...
 fack = <@\textcolor{red}{tcp\_highest\_sack\_seq}@>(<@\textcolor{mygreen}{tp}@>);
 if (!before(low_seq, fack))
  return;
 <@\textcolor{mygreen}{tp->reordering}@> = min_t(... , ...);
 <@\textcolor{mygreen}{tp->reord\_seen++;}@>
 ...
}

static inline u32 tcp_highest_sack_seq(struct tcp_sock *<@\textcolor{mygreen}{tp}@>)
{
 if (!<@\textcolor{mygreen}{tp->sacked\_out}@>)
  return tp->snd_una;
 if (<@\textcolor{mygreen}{tp->highest\_sack}@> == NULL)
  return tp->snd_nxt;
 return TCP_SKB_CB(tp->highest_sack)->seq;
}
        \end{lstlisting}
    \end{minipage}
    \caption{Code snippet of a constrained value write.}
    \label{fig:Code snippet of a constrained value write}
\end{figure}

In this section, we provide a few case studies to highlight the process through \projName{} successfully converted 
low-risk bugs into high-risk ones.

\textbf{OOB read $\Longrightarrow$ Arbitrary address write.}
Figure \ref{fig:Code snippet of a arbitrary address write} shows a real case on syzbot
titled ``KASAN: slab-out-of-bounds Read in find\_first\_zero\_bit''~\cite{case_study_1}.
Starting at line 3, \texttt{info->si\_imap} was already out-of-bounds (we omit the code earlier), 
which is passed as an argument to \texttt{find\_first\_zero\_bit()}.
Then, there is an OOB read impact caught by \texttt{KASAN} at line 16.
However, if line 17 is executed next, which does occur during fuzzing, it will lead to a write impact in \texttt{set\_bit()} (invoked at at line 8).
Since we allow the kernel to continue executing despite a bug impact being caught by \texttt{KASAN} (see \S\ref{sec:impl_fuzzing}),
the write impact is reported during the fuzzing process.
Specifically, \texttt{set\_bit()} will take the \texttt{info->si\_imap} as a memory address and set one bit to \texttt{ino}.
During the symbolic execution (the scope of which is guided by static analysis), 
we can quickly determine the write impact is feasible and determine that it is an arbitrary write primitive
because the only constraint on \texttt{info->si\_imap} is \texttt{addr[idx] != ~0UL} at line 16.

\textbf{UAF read $\Longrightarrow$ UAF write.}
Figure \ref{fig:Code snippet of a constrained value write} is a real example from a use-after-free read bug~\cite{case_study_2} on syzbot, \texttt{tp} is a freed object, KASAN catches the UAF read at line 15.
There are two UAF write impacts at line 8 and 9 afterwards.
Due to the non-zero value of \texttt{tp->sacked\_out} at line 15 and \texttt{tp->highest\_sack} at line 17,
\texttt{tcp\_highest\_sack\_seq()} consistently return at line 19, which makes the following condition at line 6 always true.
This prevents the two write impacts from being unreachable.
However, since we know the entire \texttt{tp} object was freed, we can in principle control the value of the object by heap spraying.
Therefore, symbolic execution will determine the right values of \texttt{tp->sacked\_out} and \texttt{tp->highest\_sack}
in order to reach the write impacts.
Note that according to \S\ref{sec:validation}, these two write impacts are considered ``UAF write primitives'' 
because it writes to the symbolic memory (\ie freed object \texttt{tp}), as opposed to the above case study 
where the we write to a symbolic address.
In order to exploit the bug further, 
we will need to spray an object that has a function pointer or data pointer which will be overwritten (due to line 8 and 9).

\begin{figure}[t]
    \centering
    \begin{minipage}[t]{0.9\linewidth}
        \begin{lstlisting}[language=C]
bool refcount_dec_and_mutex_lock(refcount_t *<@\textcolor{mygreen}{r}@>, struct mutex *<@\textcolor{mygreen}{lock}@>)
{
 if (refcount_dec_not_one(<@\textcolor{mygreen}{r}@>))
  return false;
 ...
 return true;
}

bool refcount_dec_not_one(refcount_t *<@\textcolor{mygreen}{r}@>)
{
 unsigned int new, <@\textcolor{mygreen}{val}@> = atomic_read(&<@\textcolor{mygreen}{r->refs}@>);
 
 do {
  if (unlikely(<@\textcolor{mygreen}{val}@> == UINT_MAX))
   return true;
  if (<@\textcolor{mygreen}{val}@> == 1)
   return false;
  <@\textcolor{mygreen}{new}@> = <@\textcolor{mygreen}{val}@> - 1;
  if (<@\textcolor{mygreen}{new}@> > <@\textcolor{mygreen}{val}@>) {
   ...
   return true;
  }
 } while (...);
 return true;
}
        \end{lstlisting}
    \end{minipage}
    
    \caption{Code snippet of refcount\_dec\_and\_mutex\_lock and refcount\_dec\_not\_one}
    \label{fig:Code snippet of refcount_dec_and_mutex_lock}
\end{figure}
\label{case study:exploit case study}

\textbf{UAF/OOB read $\Longrightarrow$ func-ptr-deref}.
Figure \ref{fig:A example case of a slab-out-of-bounds Read bug} illustrates a real case~\cite{case_study_3} on syzbot 
initially marked as an OOB read. It was described as the motivating example already.
To exploit the bug, we need to reach line 36 to dereference a function pointer from the out-of-bounds object or its derived objects, and 
then we can hijack the control flow.
Again, due to the fact that the fuzzing process is blocked at line 14 due to an exception, 
it required symbolic execution to provide a legitimate value of \texttt{actions}.
In addition, at line 28, there is another condition regarding the return value of 
\texttt{refcount\_dec\_and\_mutex\_lock()} (invoked at line 28), which has to turn true.
Figure \ref{fig:Code snippet of refcount_dec_and_mutex_lock} illustrates the internals of the function.
Here, the parameters \texttt{r} and \texttt{lock} are still from out-of-bounds memory, and therefore take symoblic values.
In order to determine the right values of them, we need to follow into \texttt{refcount\_dec\_not\_one()}
and make sure that it will return false, in order to return true at line 6 ultimately, we have determined that \texttt{val} should be 1 in order for this to happen.
As we can see, this example is more convoluted involving more conditions and more path exploration by symbolic execution.
As a result, the symbolic execution is guided according to the direction at each condition, which eventually finishes the exploration
in only 18 seconds. Without guidance from static analysis, symbolic execution is unable to find this function pointer dereference in 2 hours.
We have managed to write an actual exploit that does the actual heap spraying and show that we can hijack the control flow, \ie 
execute code at an arbitrary address.

\subsection{Disclosure}

In this section, we validate the utility of \projName{} in terms of helping with timely patch propagation to downstream kernels (for fixed bugs), as well as enabling the prioritization of bug fixes in the upstream (for open bugs).

\vspace{0.014in}
\noindent\textbf{Downstream kernels.} 
We initially attempted to report to downstream kernels (\eg Ubuntu) and have them apply patches from upstream after we determine that the patches are missing. This is unfortunately time-consuming as we have to manually analyze the source to determine which downstream kernels are actually affected by the bug. Eventually, we followed the suggestion from a Ubuntu kernel developer~\cite{discuss_with_ubuntu} and resort to the CVE Numbering Authority. Interestingly, we observe that the majority of the syzbot bugs do not have any CVEs assigned to them. Thus, if we are able to successfully request CVEs for the high-risk bugs, we can potentially benefit all the distributions that follow the CVE database.
We reported all 32 high-risk bugs after Linux kernel v4.19. At the time of writing, 8 of them have already been assigned CVEs and we have not heard back from the remaining 24.
As an example demonstrating the effectiveness of this strategy, the publication of CVE-2021-33034 has led to immediate actions by Ubuntu~\cite{Ubuntu_cve_2021-33034}, Fedora~\cite{fedora_cve_2021-33034}, RedHat~\cite{Red_Hat_cve_2021-33034}, and Debian~\cite{Debian_cve_2021-33034}.

\vspace{0.014in}
\noindent\textbf{Upstream kernels.} Since the CVE assignment is predicated on the availability of patches, 
we had to report our findings directly to upstream kernel developers. 
This process is more tricky than we anticipated. 
In total, we have reported 6 bugs in the open section out of the total 34 high-risk bugs determined by SyzScope. 
We reported a subset because it turns out that most open bugs already have a pending fix (but need time to be confirmed effective). At the time of writing, we have seen two patches submitted because of our reporting (with email replies to the same thread), demonstrating the positive influence on getting important security bugs fixed.
During the process of reporting, while we have received appreciation of our project and reports, 
we have also learned that there is a serious lack of resources to fix syzbot-generated bugs,
which is likely the reason why we have not received responses for the other 4 bugs.
Nevertheless, we do receive some positive feedback on our research~\cite{SyzScope_debate_positive}.

\section{Discussion}
\label{sec:discussion}

\vspace{0.018in}
\noindent\textbf{Evaluating bugs that are already high-impact.}
In this project, we take only low-risk bugs as input to \projName{}. However, the result shows the number of high-risk impacts associated
with a single bug can be extremely high. Specifically, among the \numOfHighRiskBugs{} high-risk bug cases, we find 4,843 high-risk impacts.
Moreover, not all high-risk impacts are equal, \eg func-ptr-deref being the most dangerous. Therefore, even if a bug already exhibits
some high-risk impact, \eg OOB write, we can still feed it to \projName{} to uncover even more high-risk impacts.
This can be beneficial if the goal is to further determine the exploitability of a bug.

\vspace{0.018in}
\noindent\textbf{Supporting more types of bug impacts.}
\projName{} currently focused on modeling OOB and UAF, starting from the hidden impact estimation component, \ie identifying the vulnerable objects and symbolizing the memory that the adversary can control.
However, there are still other bug types such as unitialized use of memory which we can support in the future. 

\vspace{0.018in}
\noindent\textbf{Interfacing with other exploitability testing systems.}
\projName{} is complementary to these projects that aim to automatically or semiautomatically
evaluate the exploitability of kernel bugs. Fuze~\cite{DBLP:conf/uss/WuCXXGZ18} and KOOBE~\cite{DBLP:conf/uss/ChenZLQ20} 
are two representative projects that target UAF and OOB bugs respectively.
For example, KOOBE~\cite{DBLP:conf/uss/ChenZLQ20} can work with only bugs that exhibit an OOB write impact already and 
ignore any OOB read bugs by design.
According to our results, \projName{} is able to turn 32 OOB read bugs into OOB write ones, which would allow KOOBE to almost double the number 
of cases that it has evaluated against. 

\vspace{0.018in}
\noindent\textbf{Pending patches for Syzbot open bugs.}
Syzbot recently added a new feature called \textit{Patch testing requests}. This feature allows developers and maintainers to upload patches for bug testing, if a patch managed to eliminate the bug, syzbot will release such patches on the dashboard and the patch will be merged into the upstream. This feature speeds up the patching process by automating the testing of patches.
We note that it is possible to use those pending patches in the context exploration process when it comes to eliminate unrelated bugs. However, we did not choose to do this because 
such patches are not officially accepted by Linux and can potentially lead to misleading results.

\section{Related Work}

\noindent\textbf{Kernel fuzzing.}
There are several prominent general-purpose kernel fuzzers that have been developed to discover security bugs~\cite{syzkaller, TriforceLinuxSyscallFuzzer, trinity}. More recently, many projects have focused on improving various aspects of kernel fuzzing~\cite{DBLP:conf/uss/PailoorAJ18, DBLP:conf/ndss/KimJKJSL20, DBLP:conf/sp/JeongKSLS19, DBLP:conf/sp/XuKZK20}.
For example, MoonShine~\cite{DBLP:conf/uss/PailoorAJ18} aims to provide highly distilled seeds to bootstrap the fuzzing process.
HFL~\cite{DBLP:conf/ndss/KimJKJSL20} proposed a hybrid fuzzing solution to address a few weaknesses encountered in coverage-guided fuzzing, 
\eg identifying explicit dependencies among syscalls.
Razzer~\cite{DBLP:conf/sp/JeongKSLS19} and KRACE~\cite{DBLP:conf/sp/XuKZK20} improve the fuzzing logic specifically against race condition bugs.
Unfortunately, these general-purpose or custom kernel fuzzers all target at uncovering more bugs quickly 
instead of understanding the security impacts of reported bugs (which are often manually investigated afterwards).
Finally, in addition to syzbot, continuous fuzzing against Linux kernels has become a common practice in the industry~\cite{DBLP:conf/sigsoft/ShiWFWSJSJS19}.

\vspace{0.018in}
\noindent\textbf{Security impact of Linux kernel bugs and crash deduplication.} 
A recent talk at Linux Security Summit by Dmitry Vyukov has shown that some seemingly low-risk bugs from syzbot turn out to be of high risks~\cite{Dmitry_Vyukov}.
However, no systematic and automated solution has been proposed to understand this phenomenon on a large scale.
A closely related work~\cite{DBLP:conf/ndss/WuHML20} aims to infer the security impact of a bug by analyzing its patches (as opposed to a bug reproducer and sanitizer-generated bug report).
Unfortunately, due to the limited information provided by a patch, they were able to identify only 227 patches that are fixing security bugs 
(with only 243 security impacts identified) after scanning 54,000 commits. 
This is in contrast with the 4,843 security impacts that we discover from only \numOfHighRiskBugs{} low-risk bugs.
We believe the incomplete result is partly due to the lack of runtime information which forces the analysis to be constrained at a local scale.
Furthermore, there is no differentiation of high-risk and low-risk impacts, which is a key goal of \projName{}.
Finally, there have been many studies on determining the security impact of bugs based on text analysis (\eg bug report descriptions) using data mining and machine learning ~\cite{mining_bug_databases,6798341,8424985,graduate_thesis_reports,vulnerability_identification}.

\vspace{0.018in}
\noindent\textbf{Exploitability testing.} 
Recent work has attempted to turn different types of Linux kernel security bugs into actual exploits in an automated or semi-automated fashion.
Fuze~\cite{DBLP:conf/uss/WuCXXGZ18} and KOOBE~\cite{DBLP:conf/uss/ChenZLQ20} are two representative projects that 
target use-after-free (UAF) and out-of-bound (OOB) bugs respectively. 
They take in a proof-of-concept (PoC) that can trigger a bug (often only crashing the kernel) as input, 
and then conduct various analyses to convert the PoC into an exploit.
In addition, there are also related work on exploiting other types of security bugs such as uninitialized uses of stack variables~\cite{stack_uninit},
which we did not support in our current implementation, due to the fact the corresponding \texttt{KMSAN} sanitizer currently does not provide sufficient details in the bug report.
Finally, there are also related work that aim to assist the process of generating an exploit~\cite{slake,236346}.
All of these related work are complementary to \projName{}, where our system can be considered a frontend that can interface with these systems as backends.
In addition to kernel exploits, there are also other related work on userspace analyzing the exploitability of heap bugs~\cite{heaphopper, auto-heap-exploit}.

\section{Conclusion}

In this paper, we perform a systematic investigation on fuzzer-exposed bugs on the syzbot platform.
In order to conduct such an analysis, we develop an automated pipeline that combines fuzzing, static analysis,
and symbolic execution together to reveal high-risk security impacts of seemingly low-risk bugs.
The system can be easily integrated into the syzbot platform that continuously evaluates newly discovered bugs.
After analyzing over one thousand such bugs, we demonstrate that \numOfBugToHighRisk{} of them can be turned into
high-risk bugs. The results have important implications on patch prioritization and propagation moving forward.
To facilitate reproduction and future research, we open source SyzScope at https://github.com/seclab-ucr/SyzScope.

\bibliographystyle{plain}
\bibliography{sample}

\begin{thebibliography}{10}

\bibitem{write-what-where}
{CWE-123: Write-what-where Condition}.
\newblock \url{https://cwe.mitre.org/data/definitions/123.html}.

\bibitem{Debian_cve_2021-33034}
Debian cve-2021-33034.
\newblock \url{https://security-tracker.debian.org/tracker/CVE-2021-33034}.

\bibitem{discuss_with_ubuntu}
Discussion with ubuntu maintainers.
\newblock
  \url{https://lists.ubuntu.com/archives/ubuntu-devel/2021-May/041457.html}.

\bibitem{fedora_cve_2021-33034}
Fedora cve-2021-33034.
\newblock
  \url{https://lists.fedoraproject.org/archives/list/package-announce@lists.fedoraproject.org/message/GI7Z7UBWBGD3ABNIL2DC7RQDCGA4UVQW/}.

\bibitem{case_study_1}
Kasan: slab-out-of-bounds read in find\_first\_zero\_bit.
\newblock
  \url{https://syzkaller.appspot.com/bug?id=93d1336016a1e1cdade11339e85f7b85ce3d4abc}.

\bibitem{case_study_3}
Kasan: slab-out-of-bounds read in tcf\_exts\_destroy.
\newblock
  \url{https://syzkaller.appspot.com/bug?id=2389bfc4b1c4ea3969629ed19bef0b3b2ec741f2}.

\bibitem{case_study_2}
Kasan: use-after-free read in tcp\_fastretrans\_alert.
\newblock
  \url{https://syzkaller.appspot.com/bug?id=7be8b464a3a27e6dc5c73d3ffe3b56dc0cf51e52}.

\bibitem{s2e_single_core}
{Parallel S2E}.
\newblock \url{http://s2e.systems/docs/Howtos/Parallel.html}.

\bibitem{Red_Hat_cve_2021-33034}
Red hat cve-2021-33034.
\newblock \url{https://access.redhat.com/security/cve/cve-2021-33034}.

\bibitem{SyzScope_debate_positive}
Syzscope debate positive opinion.
\newblock \url{https://lore.kernel.org/lkml/YL3zGGMRwmD7fNK+@zx2c4.com/}.

\bibitem{SyzScope_github_repo}
Syzscope github repo.
\newblock \url{https://github.com/seclab-ucr/SyzScope/}.

\bibitem{Ubuntu_cve_2021-33034}
Ubuntu cve-2021-33034.
\newblock \url{https://ubuntu.com/security/CVE-2021-33034}.

\bibitem{UBSAN}
The undefined behavior sanitizer - ubsan.
\newblock \url{https://www.kernel.org/doc/html/v4.14/dev-tools/ubsan.html}.

\bibitem{syzkaller_intro}
{Coverage-guided kernel fuzzing with syzkaller}.
\newblock \url{https://lwn.net/Articles/677764/}, 2016.

\bibitem{angr}
a binary analysis framework.
\newblock \url{https://angr.io/}, 2017.

\bibitem{fix_cve-2019-2215}
{ANDROID: binder: remove waitqueue when thread exits}.
\newblock
  \url{https://git.kernel.org/pub/scm/linux/kernel/git/torvalds/linux.git/commit/?id=f5cb779ba16334b45ba8946d6bfa6d9834d1527f},
  2018.

\bibitem{cve-2019-2215}
{CVE-2019-2215 - Bad Binder: Android In-The-Wild Exploit}.
\newblock
  \url{https://googleprojectzero.blogspot.com/2019/11/bad-binder-android-in-wild-exploit.html},
  2019.

\bibitem{KASAN}
Kernel addresssanitizer.
\newblock \url{https://github.com/google/kasan}, 2020.

\bibitem{6798341}
D.~{Behl}, S.~{Handa}, and A.~{Arora}.
\newblock A bug mining tool to identify and analyze security bugs using naive
  bayes and tf-idf.
\newblock In {\em 2014 International Conference on Reliability Optimization and
  Information Technology (ICROIT)}, 2014.

\bibitem{DBLP:conf/uss/ChenZLQ20}
Weiteng Chen, Xiaochen Zou, Guoren Li, and Zhiyun Qian.
\newblock {KOOBE:} towards facilitating exploit generation of kernel
  out-of-bounds write vulnerabilities.
\newblock In {\em 29th {USENIX} Security Symposium}, 2020.

\bibitem{slake}
Yueqi Chen and Xinyu Xing.
\newblock Slake: Facilitating slab manipulation for exploiting vulnerabilities
  in the linux kernel.
\newblock In {\em Proceedings of the 2019 ACM SIGSAC Conference on Computer and
  Communications Security (CCS)}, 2019.

\bibitem{s2e}
Vitaly Chipounov, Volodymyr Kuznetsov, and George Candea.
\newblock S2e: A platform for in-vivo multi-path analysis of software systems.
\newblock 2011.

\bibitem{PT-Rand}
Lucas Davi, David Gens, Christopher Liebchen, and Ahmad-Reza Sadeghi.
\newblock Pt-rand: Practical mitigation of data-only attacks against page
  tables.
\newblock In {\em Proc. of 24th Annual Network \& Distributed System Security
  Symposium (NDSS)}, 2017.

\bibitem{heaphopper}
Moritz Eckert, Antonio Bianchi, Ruoyu Wang, Yan Shoshitaishvili, Christopher
  Kruegel, and Giovanni Vigna.
\newblock Heaphopper: Bringing bounded model checking to heap implementation
  security.
\newblock In {\em 27th {USENIX} Security Symposium ({USENIX} Security 18)},
  pages 99--116, Baltimore, MD, August 2018. {USENIX} Association.

\bibitem{oss-fuzz}
Google.
\newblock Oss-fuzz.
\newblock \url{https://google.github.io/oss-fuzz/}, 2016.

\bibitem{KCSAN}
Google.
\newblock Kcsan.
\newblock \url{https://www.kernel.org/doc/html/latest/dev-tools/kcsan.html},
  2020.

\bibitem{KMSAN}
Google.
\newblock Kmsan.
\newblock \url{https://github.com/google/kmsan}, 2020.

\bibitem{syzbot}
{Google}.
\newblock {syzbot}.
\newblock \url{https://syzkaller.appspot.com/upstream/}, 2020.

\bibitem{syzkaller}
{Google}.
\newblock {syzkaller}.
\newblock \url{https://github.com/google/syzkaller}, 2020.

\bibitem{8424985}
K.~{Goseva-Popstojanova} and J.~{Tyo}.
\newblock Identification of security related bug reports via text mining using
  supervised and unsupervised classification.
\newblock In {\em 2018 IEEE International Conference on Software Quality,
  Reliability and Security (QRS)}, 2018.

\bibitem{androidbinder}
Hongli Han and Mingjian Zhou.
\newblock Android binder: The bridge to root.
\newblock
  \url{https://conference.hitb.org/hitbsecconf2019ams/sessions/binder-the-bridge-to-root/},
  2019.

\bibitem{cve-2019-2025}
Hongli Han and Mingjian Zhou.
\newblock {CVE-2019-2025 - Android Binder: The Bridge To Root}.
\newblock
  \url{https://conference.hitb.org/hitbsecconf2019ams/materials/D2T2%20-%20Binder%20-%20The%20Bridge%20to%20Root%20-%20Hongli%20Han%20&%20Mingjian%20Zhou.pdf},
  2019.
\newblock HITBSecConf.

\bibitem{DBLP:conf/sp/JeongKSLS19}
Dae~R. Jeong, Kyungtae Kim, Basavesh Shivakumar, Byoungyoung Lee, and Insik
  Shin.
\newblock Razzer: Finding kernel race bugs through fuzzing.
\newblock In {\em 40th {IEEE} Symposium on Security and Privacy}, 2019.

\bibitem{trinity}
Dave Jones.
\newblock trinity.
\newblock \url{https://github.com/kernelslacker/trinity}, 2011.

\bibitem{TriforceLinuxSyscallFuzzer}
Dave Jones.
\newblock Triforce linux syscall fuzzer.
\newblock \url{https://github.com/nccgroup/TriforceLinuxSyscallFuzzer}, 2016.

\bibitem{DBLP:conf/ndss/KimJKJSL20}
Kyungtae Kim, Dae~R. Jeong, Chung~Hwan Kim, Yeongjin Jang, Insik Shin, and
  Byoungyoung Lee.
\newblock {HFL:} hybrid fuzzing on the linux kernel.
\newblock In {\em 27th Annual Network and Distributed System Security
  Symposium, {NDSS}}, 2020.

\bibitem{stack_uninit}
Kangjie Lu, Marie-Therese Walter, David Pfaff, Stefan Nürnberger, Wenke Lee,
  and Michael Backes.
\newblock Unleashing use-before-initialization vulnerabilities in the linux
  kernel using targeted stack spraying.
\newblock In {\em 27th Annual Network and Distributed System Security
  Symposium, {NDSS}}, 2017.

\bibitem{DBLP:conf/uss/MachirySCSKV17}
Aravind Machiry, Chad Spensky, Jake Corina, Nick Stephens, Christopher Kruegel,
  and Giovanni Vigna.
\newblock {DR.} {CHECKER:} {A} soundy analysis for linux kernel drivers.
\newblock In {\em 26th {USENIX} Security Symposium}, 2017.

\bibitem{DBLP:conf/uss/PailoorAJ18}
Shankara Pailoor, Andrew Aday, and Suman Jana.
\newblock {MoonShine:} optimizing os fuzzer seed selection with trace
  distillation.
\newblock In {\em 27th {USENIX} Security Symposium}, 2018.

\bibitem{DBLP:conf/sigsoft/ShiWFWSJSJS19}
Heyuan Shi, Runzhe Wang, Ying Fu, Mingzhe Wang, Xiaohai Shi, Xun Jiao, Houbing
  Song, Yu~Jiang, and Jiaguang Sun.
\newblock Industry practice of coverage-guided enterprise linux kernel fuzzing.
\newblock In {\em ESEC/FSE}, 2019.

\bibitem{graduate_thesis_reports}
Jacob~P. Tyo.
\newblock Empirical analysis and automated classification of security bug
  reports.
\newblock In {\em Graduate Thesis and Reports}, 2016.

\bibitem{Dmitry_Vyukov}
Dmitry Vyukov.
\newblock Syzbot and the tale of thousand kernel bugs - dmitry vyukov, google.
\newblock \url{https://youtu.be/qrBVXxZDVQY}, 2018.

\bibitem{UBSAN_enable}
Dmitry Vyukov.
\newblock syzbot: enable ubsan for linux.
\newblock \url{https://github.com/google/syzkaller/issues/1523}, 2020.

\bibitem{cve-2018-9568}
Yong Wang.
\newblock {CVE-2018-9568 - From Zero to Root: Building Universal Android
  Rooting With a Type Confusion Vulnerability}.
\newblock
  \url{https://github.com/ThomasKing2014/slides/blob/master/Building%20universal%20Android%20rooting%20with%20a%20type%20confusion%20vulnerability.pdf},
  2019.
\newblock Zer0Con.

\bibitem{vulnerability_identification}
Dumidu Wijayasekara, Milos Manic, and Miles McQueen.
\newblock Vulnerability identification and classification via text mining bug
  databases.
\newblock In {\em IECON 2014 - 40th Annual Conference of the IEEE Industrial
  Electronics Society}, 2014.

\bibitem{mining_bug_databases}
Dumidu Wijayasekara, Milos Manic, Jason~L. Wright, and Miles McQueen.
\newblock Mining bug databases for unidentified software vulnerabilities.
\newblock In {\em 5th International Conference on Human System Interactions},
  2012.

\bibitem{DBLP:conf/ndss/WuHML20}
Qiushi Wu, Yang He, Stephen McCamant, and Kangjie Lu.
\newblock Precisely characterizing security impact in a flood of patches via
  symbolic rule comparison.
\newblock In {\em 27th Annual Network and Distributed System Security
  Symposium, {NDSS}}, 2020.

\bibitem{236346}
Wei Wu, Yueqi Chen, Xinyu Xing, and Wei Zou.
\newblock {KEPLER}: Facilitating control-flow hijacking primitive evaluation
  for linux kernel vulnerabilities.
\newblock In {\em 28th {USENIX} Security Symposium ({USENIX} Security 19)},
  pages 1187--1204, Santa Clara, CA, August 2019. {USENIX} Association.

\bibitem{DBLP:conf/uss/WuCXXGZ18}
Wei Wu, Yueqi Chen, Jun Xu, Xinyu Xing, Xiaorui Gong, and Wei Zou.
\newblock {FUZE:} towards facilitating exploit generation for kernel
  use-after-free vulnerabilities.
\newblock In {\em 27th {USENIX} Security Symposium, {USENIX} Security}, 2018.

\bibitem{DBLP:conf/sp/XuKZK20}
Meng Xu, Sanidhya Kashyap, Hanqing Zhao, and Taesoo Kim.
\newblock Krace: Data race fuzzing for kernel file systems.
\newblock In {\em 41st {IEEE} Symposium on Security and Privacy}, 2020.

\bibitem{auto-heap-exploit}
Insu Yun, Dhaval Kapil, and Taesoo Kim.
\newblock Automatic techniques to systematically discover new heap exploitation
  primitives.
\newblock In {\em 29th {USENIX} Security Symposium ({USENIX} Security 20)},
  pages 1111--1128. {USENIX} Association, August 2020.

\bibitem{E-Fiber}
Zheng Zhang, Hang Zhang, Zhiyun Qian, and Billy Lau.
\newblock An investigation of the android kernel patch ecosystem.
\newblock In {\em 30th {USENIX} Security Symposium}, 2021.

\end{thebibliography}

\end{document}